\documentstyle[12pt,epsfig]{article}
\begin{document}


\title{
Study of correlation of production and decay planes \\ in
$\pi\to 3\pi$ diffractive dissociation process on nuclei
\thanks{Supported by RFBR Grants No 98-02-16508 and 96-02-17631.}
}
\author{
A.V.Efremov, Yu.I.Ivanshin, L.G.Tkatchev, R.Ya.Zulkarneev \\
{\it  Joint Institute for Nuclear Research}  \\
{\it Dubna, 141980 Russia}  \\
}
\date{}
\maketitle
\vspace{-80mm}
{~\hfill \parbox{60mm}{Preprint JINR E1-98-371\\
             Submitted to Yad. Fiz.\\
             nucl-th/9901005}}

\vspace{+60mm}

\begin{abstract}
A large correlation of production and decay planes of
($\pi^-\pi^+\pi^-$)-system in dissociation of $\pi^-$--beam
$40\,GeV/c$ on nuclear targets was observed.
The dependence of the correlation
on atomic number, Feynmann variable and transversal momentum as well as
on invariant mass of the pion triple and neutral pion pair were
investigated. It was shown that the phenomenon has a clear dynamic origin
and resembles the single spin asymmetry behavior.
\end{abstract}


\section{Introduction}
The measurement of polarization at high energies is a complicated problem
since the observation of a spin or total angular momentum projection is a
non-direct one as a rule and explore features such as angular
distribution in secondary scattering or in decay process.  For a strong
interaction process parity and Lorentz invariance requires that at least
three momenta of particles (either spinless or spin-averaged) in a final
state were measured.

Some years ago the concept of handedness was introduced\footnote{ In fact,
an idea similar to the
handedness was earlier proposed in works \cite{early}. Its application to
certain heavy quark decays was studied in Ref. \cite{dalitz}. Similar
technique was also studied in work \cite{collins}.} as a measure
of polarization of parent partons (or decaying hadrons) \cite{hand}.
It was defined as an asymmetry of a process probability $W$ with respect to
a spatial component of an axial 4-vector $n_\mu\propto
\epsilon_{\mu\nu\sigma\rho}k_1^\nu k_2^\sigma k^\rho$, where $k^\rho$ is
4-momentum of particle  (or a system) in question ($k=k_1+k_2+k_3+\cdots$),
with respect to some direction $\mathbf{i}$ ($n_i= \mathbf{n}\mathbf{i}$)
\begin{equation}
H_i={W(n_i>0)- W(n_i<0)\over W(n_i>0) + W(n_i<0)}= \alpha_iP_i,
\label{hand}
\end{equation}
which was shown to be proportional to polarization $P_i$
(at least for spin 1/2 and spin 1),
provided the analyzing power $\alpha_i$ is not zero. The direction $\mathbf{i}$
could be chosen as longitudinal ($L$) with respect to the momentum $\mathbf{k}$
and as transversal ones ($T1$ or $T2$).

In the previous publication \cite{prev} the attention was drawn to the fact
that in diffractive production of pion triples \cite{mis}
\begin{equation}
\pi^- +A \to (\pi^- \pi^+ \pi^-)+A,
\label{react}
\end{equation}
by $\pi^-$ beam $40\ GeV/c$ from a nucleus $A$, a noticeable asymmetry with
respect to the triple production plane (transversal handedness $H_{T1}$)
was observed.  This paper is devoted to further experimental investigation
of this phenomenon. It includes a new information on the dependence of the
transversal handedness on the variables:
\begin{itemize}
\vspace{-1.8mm}
\item Atomic number of the target,
\vspace{-1.8mm}
\item Transversal momenta of the pion triple,
\vspace{-1.8mm}
\item Feynmann variable $X_F$ of the leading $\pi^-$,
\vspace{-1.8mm}
\item Invariant mass of the triple,
\vspace{-1.8mm}
\item Invariant mass of neutral pairs $\pi^+ \pi^-$.
\end{itemize}
Also the statistics was considerably increased.

\section{Definitions and notation}
For reaction (\ref{react}), let us define the normal to the plane of
production of a secondary pion triple $(\pi_f^- \pi^+ \pi_s^-)$
\begin{equation}
\mathbf{N} = ({\mathbf{v}}_{3\pi}\times {\mathbf{v}}_b)
\end{equation}
where ${\mathbf{v}}_b=\mathbf{k}_b/\epsilon_b$ and
${\mathbf{v}}_{3\pi}={\mathbf{k}}_{3\pi}/\epsilon_{3\pi}$ are
velocities of the initial $\pi^-$ beam and the center of mass of the triple
in Lab. r.f. and indices $f$ and $s$ label fast and slow $\pi^-$'s.  The
normal to the "decay plane" of the triple in its center of mass is defined as
\begin{equation}
\mathbf{n} = ({\mathbf{v}}^-_f-{\mathbf{v}}^+)\times
({\mathbf{v}}^-_s-{\mathbf{v}}^+)
\label{norm}
\end{equation}
where  ${\mathbf{v}}^-_{f(s)}$ or ${\mathbf{v}}^+$ are velocities of the fast
(slow) $\pi^-$ or $\pi^+$.

The transversal handedness according to (\ref{hand}) is\footnote{It is easy to
show that this quantity is in fact Lorentz-invariant.}
\begin{equation}
H_{T1}={W({\mathbf{N}\mathbf{n}}>0)- W({\mathbf{N}\mathbf{n}}<0)\over
W({\mathbf{N}\mathbf{n}}>0) + W({\mathbf{N}\mathbf{n}}<0)}.
\label{trahand}
\end{equation}
Two other components of the handedness connected with ${\mathbf{n}}\cdot
{\mathbf{v}}_{3\pi}$ and
${\mathbf{n}}\cdot({\mathbf{v}}_{3\pi}\times\mathbf{N})$ are forbidden by
the parity conservation in the strong interaction.

\section{Experimental results and discussion}
In this work the experimental material of Bologna--Dubna--Milan
Collaboration for diffraction production of $40\,GeV/c$ $\pi^-$ into three
pions was used.  The details of the experiment were presented in the works
 \cite{mis}. Notice here that the admixture of non-diffractive events in the
used set of experimental data was less than 1\%.

The transversal handedness (\ref{trahand}) was measured for a wide sample
of nuclear targets: $Be,\ ^{12}C$, $^{28}Si$, $^{48}Ti,\ ^{63}Cu,\
^{107}Ag,\ ^{181}Ta$ and $^{207}Pb$. The total number of selected events of
pion triples with leading $\pi^-$ was about 250,000.

The dependence of $H_{T1}$ on the atomic number $A$ is presented in Fig.1.
One can see that the handedness systematically decreases with increasing
$A$, which resembles a depolarization effect in multiple
scattering. An argument in this favor is the decrease of the effect as,
approximately, inverse nuclei radius.

The value of the asymmetry (\ref{trahand}), averaged over all nuclei is
\begin{equation}
H_{T1}=(5.96\pm0.21)\%
\label{result}
\end{equation}
Statistically, this is highly reliable verification of the existence of
correlation of the triple production and decay planes in process
(\ref{react}).

The values of two other asymmetries with respect to correlations
${\mathbf{n}}\cdot {\mathbf{v}}_{3\pi}$ and
$\mathbf{n}\cdot(\mathbf{v}_{3\pi}\times\mathbf{N})$ was found to be
comparable to zero from the same statistical material:
$H_{L}=(0.25\pm0.21)\%$ and $H_{T2}=(0.43\pm0.21)\%$ respectively. This is
by no means surprising, since they are forbidden by the parity conservation
in process (\ref{react}). Also they show  the order of magnitude of
systematic errors.

A natural question is to what extent the effect observed is due to the
kinematics or apparatus influence, in particular, due to acceptance of the
experimental setup where the events have been registered.  For this aim the
Monte-Carlo events of reaction (\ref{react}) were generated with a constant
mass spectrum of the $3\pi$--system in the interval 0.6--2.5 $GeV/c^2$ and
with the exponential decrease of the cross section in $t'=t-t_{min}$  with
the slope (for beryllium) $40\, (GeV/c)^{-2}$ found experimentally.  This
events were traced through the apparatus simulation with the same trigger
conditions as in \cite{mis} and the same selection of events and show no
transversal handedness $H_{T1}$
\begin{equation}
H_{T1}^{\rm MC}=(0.20\pm0.28)\%
\end{equation}
For two other asymmetries, forbidden by the parity conservation, the result
was $(0.00\pm0.28)\%$ and $(-0.14\pm0.28)\%$, respectively. Thus, the
effect (\ref{result}) cannot be explained by the kinematics or apparatus
influence.

To understand the nature of the effect observed, the dependence of the
handedness (\ref{trahand}) on the Feynmann variable $X_F=k_f/k_b$ of the
leading $\pi^-$, on the invariant mass of the triple $m_{3\pi}$ and its
neutral subsystem $m_{\pi^+\pi^-}$ and on the triple transversal momentum
$k_T$ was studied. From Fig.2 one can see that the handedness
(\ref{trahand}) increases with $X_F$, which resembles the behavior of the
single spin asymmetry (e.g. the pion asymmetry or the
$\Lambda$-polarization \cite{polar}).

The dependence of $H_{T1}$ on the triple invariant mass (Fig.3a) is
especially interesting. It clearly indicates two different sources of
$H_{T1}$ with comparable contributions: a resonant and a non-resonant one.
The resonance contribution is clearly seen at the mass of $a_1(1260)$ and
$\pi_2(1670)$ region and by all means is due to a non-zero polarization of
the resonances. The non-resonant background could also be polarized,
provided that the $3\pi$ system is predominantly in a state with the total
angular momentum $J\not=0$, e.g. if a neutral pair $m_{\pi^+\pi^-}$ was
predominantly produced from $\rho$-decay. Some indication of this can be
seen from Fig.3b. In this context, the growth of $H_{T1}$ in the region of
small $m_{3\pi}$, i.e. in the region of small relative momenta of pions,
looks quite intriguing.

A complicated picture of the $k_T$-dependence with a sharp deep at
$k_T=0.05$--$0.07\,GeV/c$ (Fig.4) reflects by all means the fact of
interference of the resonant and non-resonant processes in the triple
production.  With further increase of $k_T$ the handedness increases which
resembles the single spin asymmetry behavior too.

To check this assumption the events with invariant  mass $m_{3\pi}$
in the $a_1$ and $\pi_2$ resonance region $1.05$--$1.80\,GeV$ were excluded
from further analysis. This however does not lead us to a definite conclusion
since for $Be$ and $C$ the deep disappears but conserves for $Si$ with some
change of its form and width. The average value of the handedness stays
at the same level 5--11\% with high statistical significance.

Notice also that in earlier study of reaction (\ref{react}) at $4.5\,GeV$
for the proton target  at the hydrogen bubble chamber no angular dependence
of the normal $\mathbf{n}$ (\ref{norm}) was found just as in the Regge pole
exchange model, which provides a reasonable description of that
experiment \cite{rpem}.

In conclusion, a rather large handedness transversal to the production
plane was definitely observed in the diffractive production of
($\pi^-\pi^+\pi^-$) triples in the $\pi^-$--beam dissociation region.  The
phenomenon has a clear dynamical origin and in some features resembles the
single spin asymmetry behavior.  For a more detailed study, a partial wave
analysis of reaction  (\ref{react}) seems necessary for determination of
different spin states contribution to the investigated effect.

\medskip
The  authors sincerely thank the participants of the Bologna--Dubna--Milan
collaboration whose data were used in the present work.

\newpage
\bigskip
\renewcommand{\abstractname}{
\begin{center}
{\large\bf Изучение корреляций плоскостей рождения и распада \\
в процессе дифракционной диссоциации $\pi\to 3\pi$ на ядрах}\\
{\bf А.В. Ефремов, Ю.И. Иваньшин, Л.Г. Ткачёв, Р.Я. Зулькарнеев} \\
{\it Объединённый институт ядерных исследований, Дубна, 141980, Россия}
\end{center}
}
\begin{abstract}
Наблюдена большая корреляция между плоскостями рождения и распада
($\pi^-\pi^+\pi^-$)-системы в процессе диссоциации $\pi^-$-мезонов
$40\,GeV/c$ на ядерных мишенях. Исследована  её зависимость  от атомного
номера, фейнмановской переменной и поперечного импульса, а также от
инвариантной массы пионной тройки и её нейтральной подсистемы. Показано,
что это явление имеет явную динамическую природу, а его поведение
напоминает поведение одиночных асимметрий.
\end{abstract}

\bigskip
{\Large\bf Figure captions}

\large
\bigskip\noindent
Fig.1. The $A$-dependence of the handedness.

\bigskip\noindent
Fig.2.~The handedness dependence on $X_F$ of the leading $\pi^-$.

\bigskip\noindent
Fig.3. The handedness dependence on $m_{3\pi}$ (a) and
$m_{\pi^+\pi^-_f}$ (b).

\bigskip\noindent
Fig.4. The $k_T$-dependence of the handedness.

\newpage
\begin{figure}
\begin{center}
\raisebox{-10mm}{
\mbox{\epsfig{figure=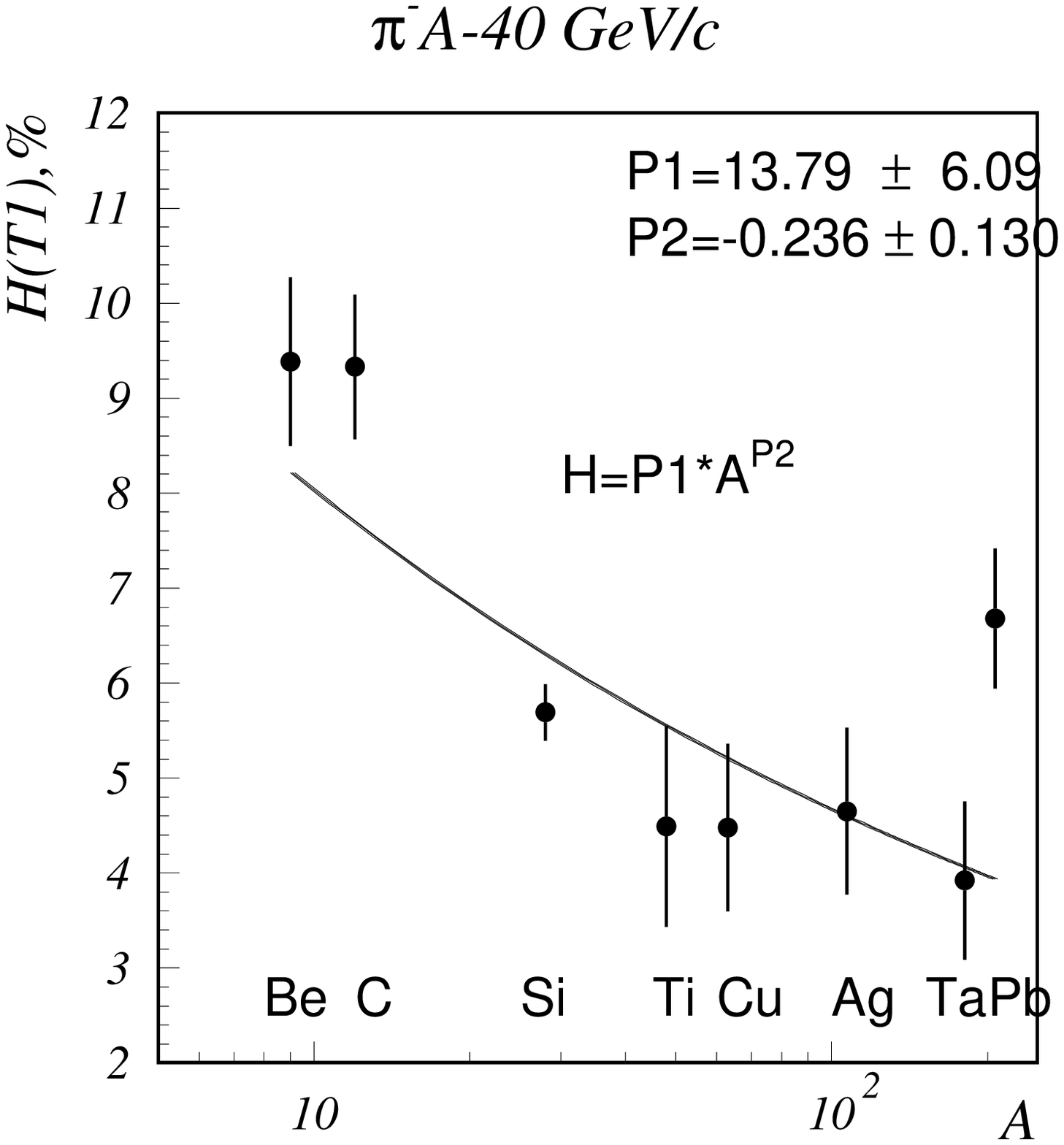,width=10.1cm,height=11cm}}}

{\footnotesize Fig.1. The $A$-dependence of the handedness.}
\end{center}
\end{figure}

\begin{figure}
\begin{center}
\raisebox{-10mm}{
\mbox{\epsfig{figure=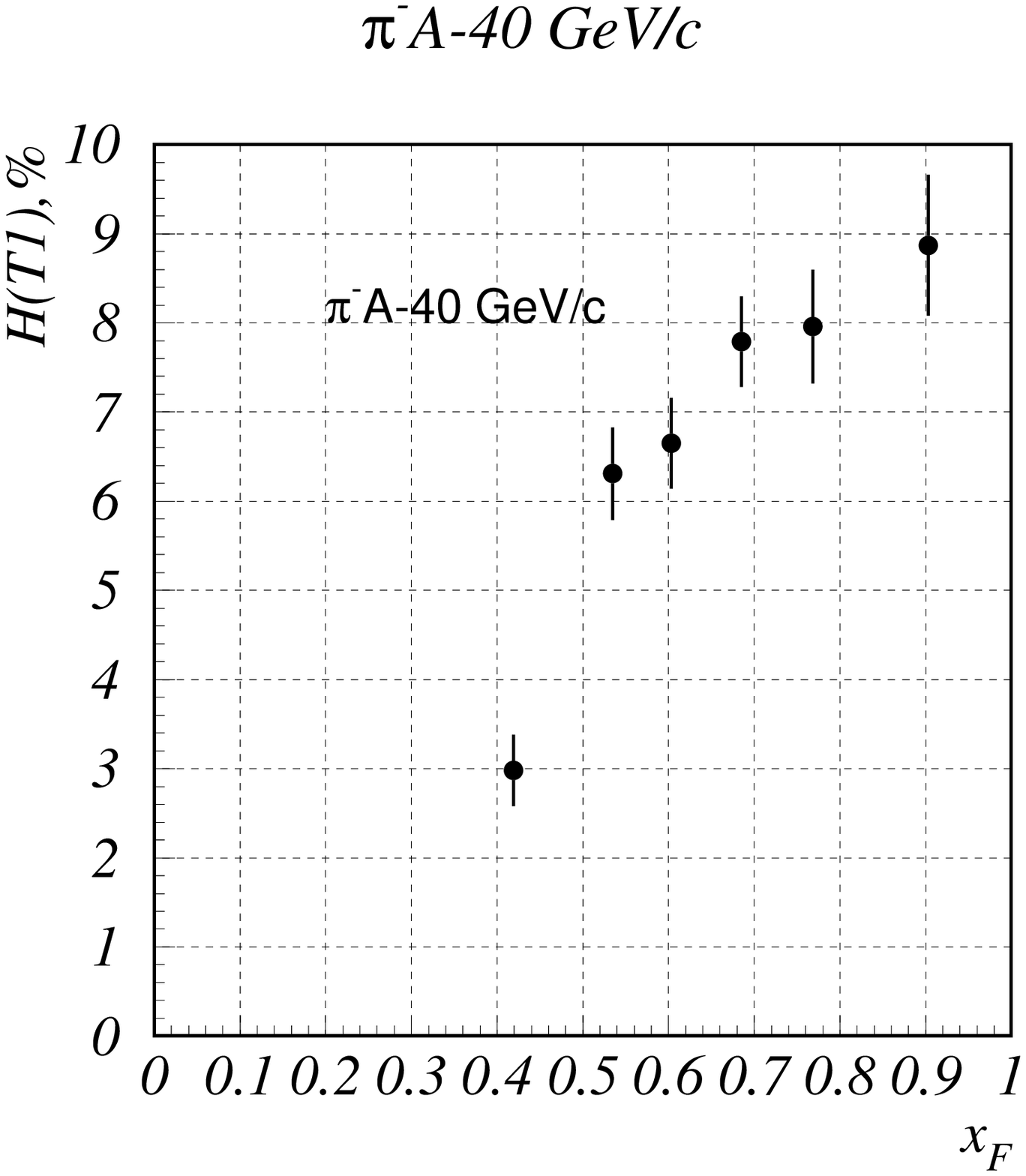,width=10.1cm,height=11cm}}}

{\footnotesize
Fig.2.~The handedness dependence on $X_F$ of the leading $\pi^-$.}
\end{center}
\end{figure}

\begin{figure}
\begin{center}
\mbox{\epsfig{figure=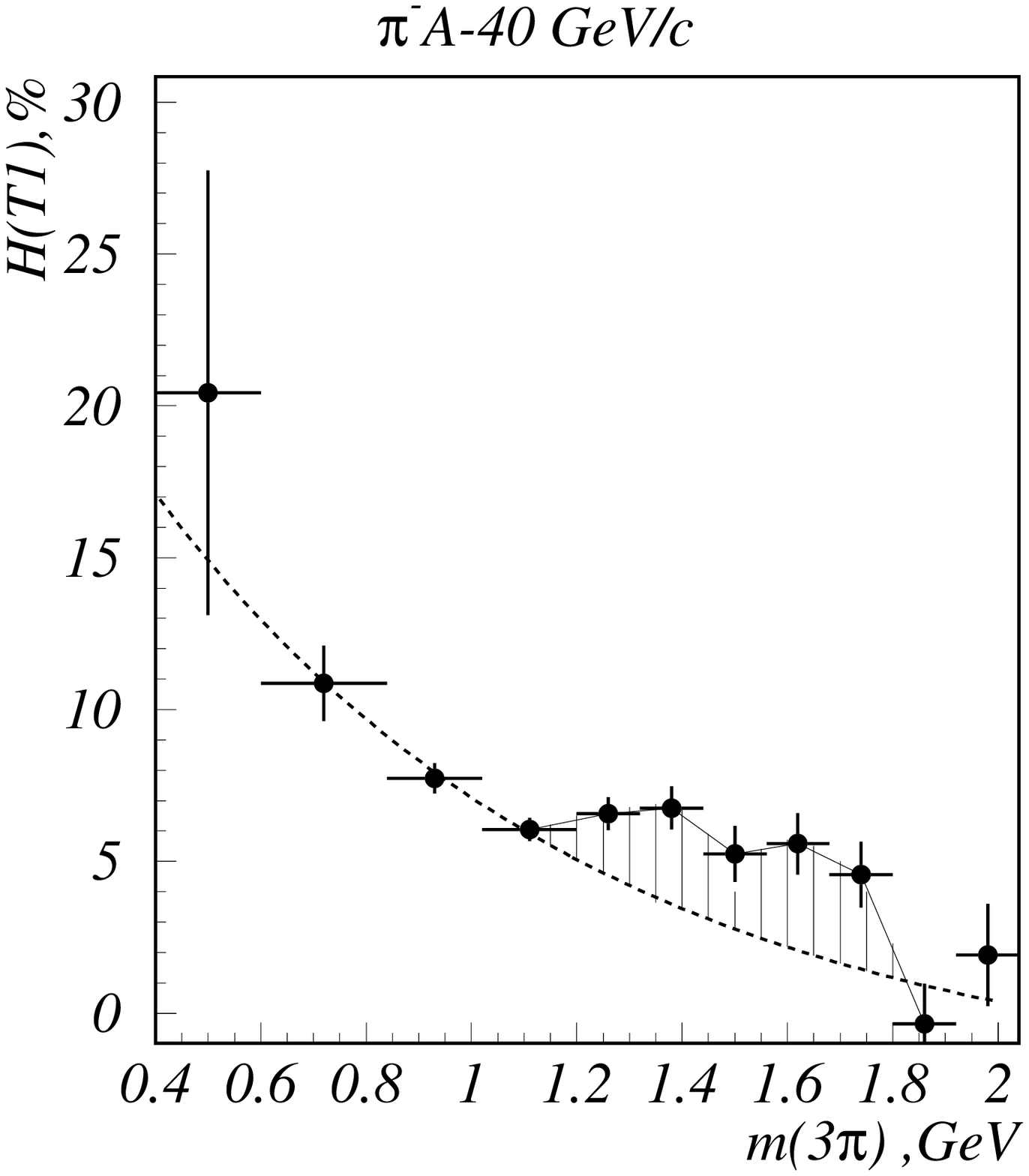,width=10.0cm,height=11cm}}
\mbox{\epsfig{figure=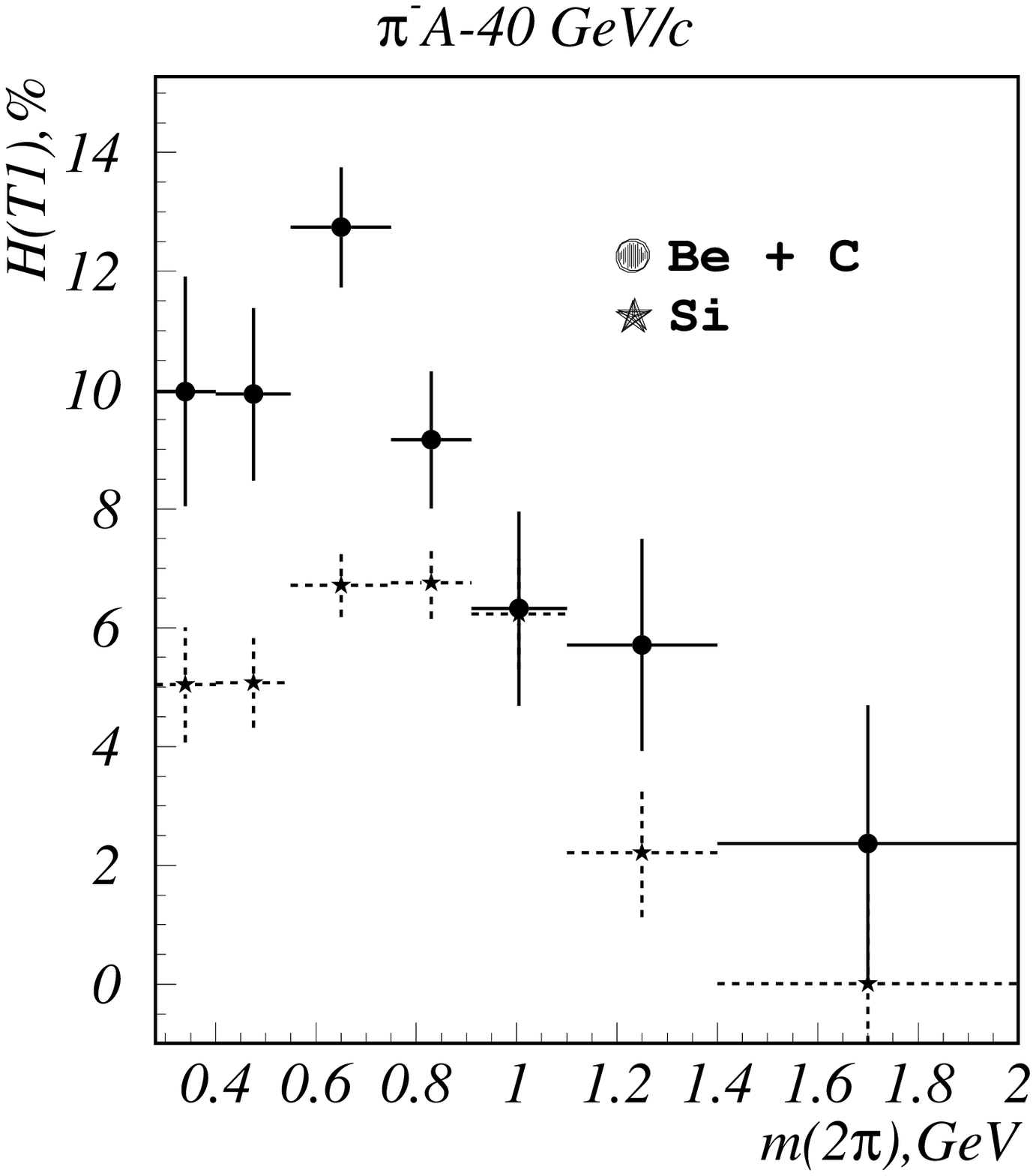,width=10.0cm,height=11cm}}\\
{\footnotesize
Fig.3. The handedness dependence on $m_{3\pi}$ (a) and
$m_{\pi^+\pi^-_f}$ (b).}
\end{center}
\end{figure}

\begin{figure}
\begin{center}
\raisebox{-10mm}{
\mbox{\epsfig{figure=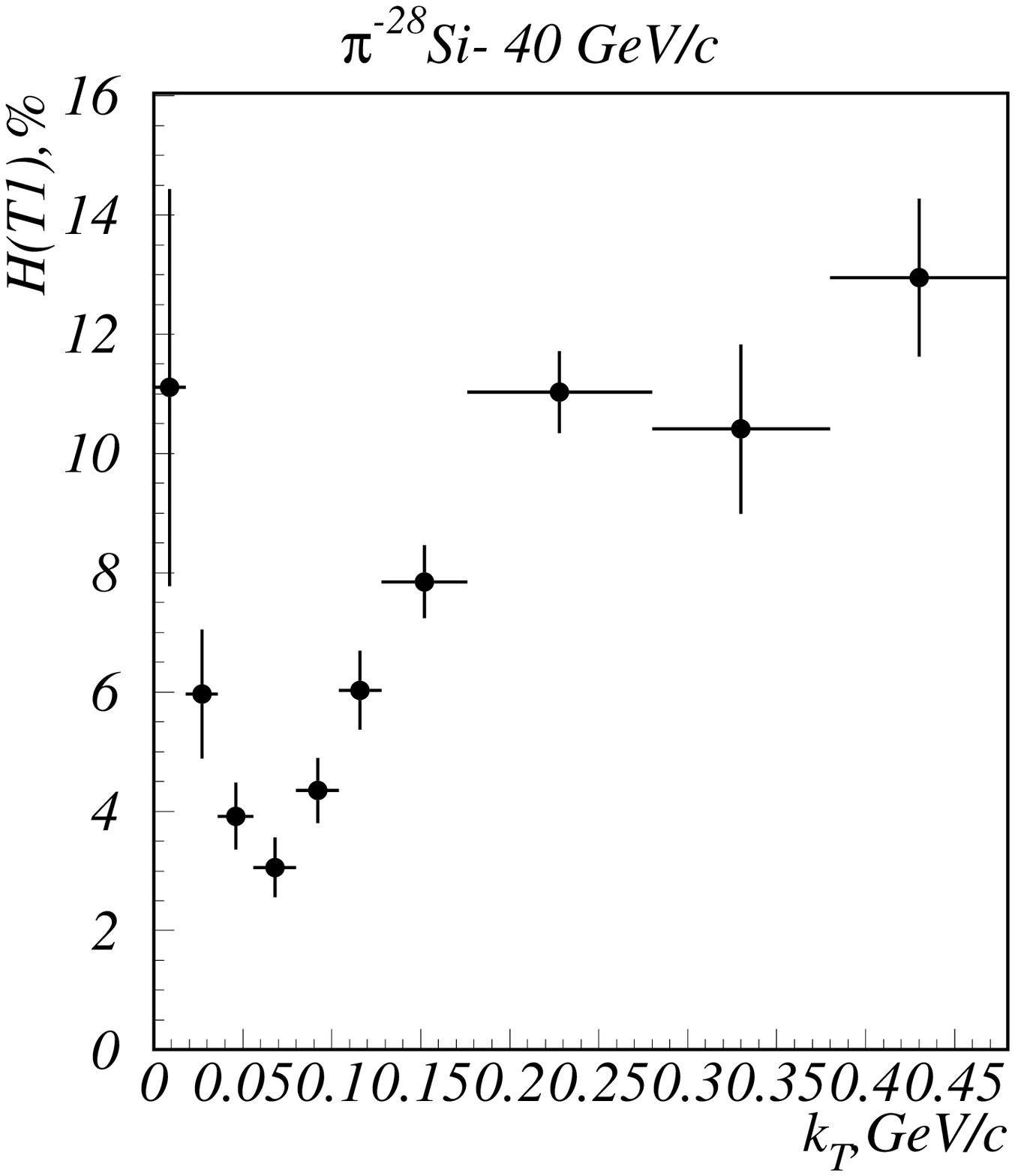,width=10.1cm,height=11.2cm}}}

{\footnotesize Fig.4. The $k_T$-dependence of the handedness.}
\end{center}
\end{figure}

\end{document}